\documentstyle[aps,preprint,tighten,floats]{revtex}
\begin{document}
\draft
\title{Gravitational waves from binary systems in circular orbits:
       Convergence of a dressed multipole expansion}
\author{Stephen W. Leonard and Eric Poisson}
\address{Department of Physics, University of Guelph, Guelph, Ontario N1G
         2W1, Canada}
\date{Submitted to Classical and Quantum Gravity, August 17, 1997}
\maketitle
\begin{abstract}
The gravitational radiation originating from a compact binary 
system in circular orbit is usually expressed as an infinite sum 
over radiative multipole moments. In a slow-motion approximation, 
each multipole moment is then expressed as a post-Newtonian 
expansion in powers of $v/c$, the ratio of the orbital velocity 
to the speed of light. The ``bare multipole truncation'' of the 
radiation consists in keeping only the leading-order (Newtonian) 
term in the post-Newtonian expansion of each moment, 
but summing over {\it all} the multipole moments. In the case of 
binary systems with small mass ratios, the bare multipole series 
was shown in a previous paper [Simone {\it et al.}~1997, 
{\it Class.~Quantum Grav.}~{\bf 14}, 237] to converge for all values 
$v/c < 2/e$, where $e$ is the base of natural logarithms. 
(These include all physically relevant values for circular inspiral.) 
In this paper, we extend the analysis to a ``dressed multipole
truncation'' of the radiation, in which the leading-order moments are 
corrected with terms of relative order $(v/c)^2$ (first post-Newtonian, 
or 1PN, terms) and $(v/c)^3$ (1.5PN terms). We find that the dressed 
multipole series converges also for all values $v/c < 2/e$, and
that it coincides (within 1 \%) with the numerically ``exact'' results 
for $v/c < 0.2$. Although the dressed multipole series converges, it is 
an unphysical approximation, and the issue of the convergence of the 
true post-Newtonian series remains uncertain. However, our analysis 
shows that an eventual failure of the true post-Newtonian series to 
converge cannot originate from summing over the Newtonian, 1PN, and 
1.5PN part of all the multipole moments.
\end{abstract}
\pacs{Pacs numbers: 0430, 0425N}
\maketitle

\section{Introduction}

Measurement of gravitational waves from inspiraling compact binaries 
(composed of neutron stars and/or black holes) by forthcoming laser 
interferometric detectors (such as LIGO \cite{LIGO} and VIRGO 
\cite{VIRGO}) will rely heavily on theoretical templates obtained 
from approximations to the exact laws of general relativity 
\cite{CutlerFlanagan,PoissonWill}.  
One possible method to calculate these templates, which has 
been pushed to a high degree of accuracy 
\cite{Blanchetetal,WillWiseman,Blanchet1}, is the post-Newtonian
approximation, which is based upon the assumption that the orbital 
motion is slow. Another approach is black-hole perturbation theory 
\cite{Tagoshietal,LeonardPoisson}, which is accurate only in the 
unrealistic situation of systems with small mass ratios. Currently, 
the post-Newtonian approach appears to be the most promising for
generating the templates. Unfortunately, however, there is 
considerable evidence that the post-Newtonian approximation has poor 
convergence \cite{Poisson,Simone}, if it converges at all. It is 
important to understand the reasons for the poor convergence. Although 
this issue is still an open one, some preliminary steps were taken in 
this direction in a previous paper \cite{paper_I}. 

The amount of energy radiated per unit time (denoted $\dot{E}$) 
by a binary system in circular orbit can be expressed as an infinite 
sum over the multipole moments of the gravitational-wave field, which 
are related to the behaviour of the source. In a slow-motion 
approximation, each multipole moment is then expressed as a 
post-Newtonian expansion in powers of $v/c$, the ratio of the 
orbital velocity to the speed of light. Since higher-order moments 
come with a higher power of the orbital velocity, the post-Newtonian 
approximation to $\dot{E}$ consists of truncating the multipole sum 
at a given multipole order, and in calculating the contributing 
moments to the appropriate post-Newtonian order. We will refer to 
this expansion of $\dot{E}$ in powers of $v/c$ as the 
``true post-Newtonian series''. At present, only a few terms 
of this post-Newtonian series have been calculated 
\cite{Blanchetetal,WillWiseman,Blanchet1}, and the issue of its 
converge is not at all clear. However, since post-Newtonian 
corrections can be computed for {\it all} the multipole moments, 
the simpler issue of the convergence of the {\it multipole} sum 
can be addressed.

In reference \cite{paper_I}, the convergence of the multipole series 
was studied in the restricted context of a ``bare multipole 
truncation'', in which the multipole moments were calculated only 
to leading-order in a post-Newtonian expansion. We refer to these 
bare moments as ``Newtonian'', 
and the bare multipole truncation therefore consists
of discarding all post-Newtonian corrections to each of the multipole 
moments. When an approximation to $\dot{E}$ is constructed by summing 
over all these bare moments, it is found that the series converges
for all orbital velocities less than a critical value. This value 
depends on the mass ratio of the system; in the case of very small
mass ratios, to which we specialize in this paper, the condition for 
convergence is $v/c < 2/e \simeq 0.7358$, where $e$ is the base 
of natural logarithms. The convergent interval includes all physically 
relevant values for the orbital velocity, which must be smaller than 
its value at the innermost stable circular orbit, $v_{\rm isco}/c = 
1/\sqrt{6} \simeq 0.4082$. Although the bare multipole truncation  
is an unphysical approximation which compares poorly with the 
numerically ``exact'' results \cite{Poisson}, this analysis reveals 
that an eventual failure of the true post-Newtonian series to converge 
cannot originate from summing over the Newtonian part of all the 
multipole moments.

In this paper, we extend the results of reference \cite{paper_I} to a 
``dressed multipole truncation'', in which the leading-order 
moments are corrected with terms of relative order $(v/c)^2$ (first 
post-Newtonian, or 1PN, terms) and $(v/c)^3$ (1.5PN terms). This improved      
analysis relies on a previous paper of ours \cite{LeonardPoisson}, in
which the equations of black-hole perturbation theory are integrated
for an arbitrary perturbing stress-energy tensor $T^{\alpha\beta}$ 
(the unperturbed spacetime is assumed to be Schwarzschild). 
Applying the results of this paper to the particular case of a particle 
moving on a circular orbit returns the desired post-Newtonian corrections 
to the bare multipole moments derived in reference \cite{PoissonSasaki}.

The dressed multipole truncation gives three separate series for 
$\dot{E}$. The first is the bare multipole series, in which we
sum over the Newtonian part of all the multipole moments. The second
and third series incorporate the 1PN and 1.5PN corrections to the
bare moments, respectively. We find that each of the series 
converges for $v/c < 2/e$, and we conclude that
inclusion of these post-Newtonian corrections does not affect the
convergence of the multipole sum. This shows that an eventual failure 
of the true post-Newtonian series to converge cannot originate from 
summing over the Newtonian, 1PN, and 1.5PN part of all the multipole 
moments. 

The following sections of the paper provide the details of the 
calculation. The dressed multipole truncation is introduced in 
Sec.~\ref{2}, and the convergence of the resulting series is 
established in Sec.~\ref{3}. Finally, concluding remarks are 
provided in Sec.~\ref{4}.

\section{Dressed multipole truncation for gravitational radiation}
\label{2}

In this section, we extend the bare multipole calculation of 
reference \cite{paper_I} by calculating the 1PN and 1.5PN corrections 
to the multipole moments of the gravitational-wave field. 
Such corrections have been calculated for general sources, first by 
Blanchet \cite{Blanchet2} using post-Newtonian theory, and then by 
us \cite{LeonardPoisson} using black-hole
perturbation theory. To carry out this calculation, we apply 
the results of reference \cite{LeonardPoisson} (hereafter referred 
to as LP) to the specific case of a point particle moving on a circular 
orbit around a Schwarzschild black hole. We denote by $\mu$ the mass of 
the particle, and by $M$ the mass of the black hole. It is assumed that
$\mu/M \ll 1$. We also denote the orbital radius by $r_0$, and the
angular velocity is $\Omega = (M/{r_0}^3)^{1/2}$. Finally, we define
the orbital velocity by $v=\Omega r_0 = (M/r_0)^{1/2}$. Throughout
we use units such that $G=c=1$.

The gravitational-wave luminosity is given by \cite{Thorne80}
\begin{equation}
\dot E = \frac{1}{32 \pi} \sum_{l=2}^{\infty} \sum_{m=-l}^{l} 
\Bigl[ \bigl| \dot{\cal I}_{lm} (u) \bigr|^2 +
\bigl| \dot{\cal S}_{lm} (u) \bigr |^2 \Bigr],    
\label{power}
\end{equation}
where ${\cal I}_{lm} (u)$ and ${\cal S}_{lm} (u)$ are the mass and 
current multipole moments, respectively, while $u$ denotes retarded 
time. (Because the orbit is fixed, $\dot{E}$ does not actually vary 
with $u$.) To evaluate $\dot{E}$, we shall use equations (5.37)--(5.39) 
of LP, which give the Fourier transform of the moments, 
$\tilde{\cal I}_{lm}(\omega)$ and $\tilde{\cal S}_{lm}(\omega)$. 
These equations incorporate post-Newtonian corrections of up to order 
$v^3$, relative to the leading-order, Newtonian expressions.

To evaluate the multipole moments, we must first compute the
source terms, which are constructed from the stress-energy tensor
of the orbiting particle \cite{PoissonI}. Using equations (5.24)--(5.27) 
of LP, we find that the only nonzero source terms are $\rho = 
R \Delta$, ${_{-1}j} = {_{-1}J} \Delta$, ${_1j} = {_1J} \Delta$, 
and ${_0t} = {_0T }\Delta$, where
\begin{eqnarray}
R &=& 1 + \frac {3v^2}{2} + O(v^4), \qquad
_{-1}J = \frac {-i v}{\sqrt{2}} \left [ 1 + \frac {3v^2}{2} + O(v^4)
\right ],  \qquad \nonumber \\
& &  \label{source} \\
_0T &=& \frac{v^2}{2} \left [ 1 + \frac {3v^2}{2} + O(v^4) \right ],
\qquad
\Delta = \frac {\mu}{r_0^2}\, \delta (r-r_0) \delta (\cos \theta) \delta
(\phi - \Omega t), \nonumber 
\end{eqnarray}
and $_1J = - \left ( _{-1}J \right )$. 

The second step is to substitute these results into equations 
(5.37)--(5.39) of LP. After a few steps of algebra, the mass and 
current multipole moments become
\begin{eqnarray}
\tilde{\cal I}_{lm} (\omega) &=& \frac {16 \pi \mu}{(2l+1)!!} 
\left [ \frac{(l+1)(l+2)}{2(l-1)l} \right ]^{1/2} 
{\mathcal{T}}_l (\omega)\, (- i \omega r_0)^l\, A_l (\omega)  
\nonumber \\ & & \mbox{}
\times (R+ 2_0T)\, Y_{lm}({\textstyle \frac{\pi}{2}},0)\, 
\delta(\omega - m \Omega),
\label{mass_1} \\
\tilde{\cal S}_{lm} (\omega) &=& 
\frac {- 16 \pi i \mu}{(2l+1)!!} \left ( \frac{l+2}{l-1} \right )^{1/2}
{\mathcal{T}}_l^{\#} (\omega)\, (- i \omega r_0)^l\, B_l (\omega)  
\nonumber \\ & & \mbox{}
\times \left [ {_{-1}J} {_{-1} Y_{lm}}({\textstyle \frac{\pi}{2}},0) 
+ {_1J} {_1 Y_{lm}} ({\textstyle \frac{\pi}{2}},0) \right ]\, 
\delta(\omega - m\Omega),
\end{eqnarray}
where
\begin{eqnarray}
A_l (\omega) &=& 1 - \frac {l+9}{2(l+1)(2l+3)} (\omega r_0)^2 - (l+2) 
\frac{M}{r} + O(v^4) , \nonumber \\
B_l (\omega) &=& 1 - \frac {l+4}{2(l+2)(2l+3)} (\omega r_0)^2 - 
\frac{(l-1)(l+2)}{l} \frac {M}{r} + O(v^4) , \label{AB} \\
| {\mathcal{T}}_l (\omega) |^2 &=& | {\mathcal{T}}_l^{\#} 
(\omega) |^2 = 1 + 2 \pi M \omega + O(v^6),   \nonumber 
\end{eqnarray}
while ${_1 Y_{lm}} (\theta,\phi)$ and ${_{-1} Y_{lm}} (\theta,\phi)$ are 
the spherical harmonics of spin weight 1 and $-1$, 
respectively \cite{spin}.

The third step is to invert the Fourier transform, which is trivial 
because of the delta function in the integrand. This yields
\begin{eqnarray} 
{\cal I}_{lm} (u) &=& \frac {16 \pi \mu}{(2l+1)!!} \left [ 
\frac{(l+1)(l+2)}{2(l-1)l} \right ]^{1/2}
{\mathcal{T}}_l (m \Omega)\, (- i m \Omega r_0)^l\, A_l (m \Omega) 
\nonumber \\ & & \mbox{}
 \times (R+ 2_0T)\, Y_{lm} ({\textstyle \frac{\pi}{2}},0)\, 
e^{- i m \Omega u} , \label {Ilm} \\
{\cal S}_{lm} (u) &=& \frac {- 16 \pi i \mu}{(2l+1)!!} 
\left ( \frac {l+2}{l-1} \right )^{1/2}
{\mathcal{T}}_l^{\#} (m \Omega)\, (- i m \Omega r_0)^l\, B_l (m \Omega)
\nonumber \\ & & \mbox{}
\times \left [ {_{-1}J} {_{-1} Y_{lm}} ({\textstyle \frac{\pi}{2}},0) + 
{_1J} {_1 Y_{lm}}({\textstyle \frac{\pi}{2}},0) \right ]\, 
e^{- i m \Omega u}.  \label{Slm}
\end{eqnarray}

It is now useful to write explicit expressions for the spin-weighted 
spherical harmonics \cite{PoissonSasaki}. First, if $l+m$ is even, then
\begin{equation}
Y_{lm} ({\textstyle \frac{\pi}{2}},0) = (-1)^{(l+m)/2} 
\left (\frac{2l+1}{4 \pi} \right )^{1/2} 
\frac{ \left [ (l-m)!(l+m)! \right]^{1/2}}{(l-m)!!(l+m)!!},
\label {Ylm}
\end{equation}
while $Y_{lm}({\textstyle \frac{\pi}{2}},0) =0$ if $l+m$ is odd. 
Second, if $l+m$ is odd, then
\begin{equation}
{_{-1}Y_{lm}} ({\textstyle \frac{\pi}{2}}, 0) = 
(-1)^{(l+m)/2} \left [ \frac {2l+1}{4 \pi l(l+1)} \right ]^{1/2} 
\frac {(l+m)!!(l-m)!!}{ \left [ (l+m)!(l-m)! \right]^{1/2}};
\label {Ylm_-1}
\end{equation}
although ${_{-1}Y_{lm}}({\textstyle \frac{\pi}{2}}, 0) \neq 0$ for 
$l+m$ even, its expression will not be needed. Finally, we note 
that ${_{1}Y_{lm}}({\textstyle \frac{\pi}{2}}, 0) = (-1)^{l+m} 
{_{-1}Y_{lm}}({\textstyle \frac{\pi}{2}}, 0)$. 

Looking at this last result and the fact that $_1 J = -(_{-1} J)$, 
we see that the quantity within the square brackets of equation 
(\ref{Slm}) may be simplified as follows: it vanishes identically 
for $l+m$ even, and it is equal to 
$2 {_{-1}J}\, {_{-1}Y_{lm}} ({\textstyle \frac{\pi}{2}},0)$ 
when $l+m$ is odd. Using this while substituting equations (\ref{Ilm}) 
and (\ref{Slm}) into equation (\ref{power}), we obtain 
\begin{equation}
\dot E = \dot E_Q \sum_{l=2}^\infty \sum_{m=-l}^{l} 
{\textstyle \frac {1}{2}} \eta_{lm} = \dot E_Q 
\sum_{l=2}^\infty \sum_{m=1}^{l} \eta_{lm}, 
\label{power_1}
\end{equation}
where
\begin{equation}
\dot E_Q = \frac {32}{5} \left (\frac {\mu}{M} \right )^2 v^{10}
\end{equation}
is the quadrupole-formula result. We also have
\begin{equation}
\eta_{lm} = \frac {5 \pi}{4} \frac {m^{2(l+1)}}{(2l+1)!!^2} v^{2(l-2)}
\left(1+2 \pi m v^3 \right)\, \zeta_{lm},
\label{eta}
\end{equation}
where
\begin{equation}
\zeta_{lm} = \frac{(l+1)(l+2)}{2(l-1)l} \left [ A_l (R+2_0T)\, 
Y_{lm} ({\textstyle \frac{\pi}{2}},0) \right ]^2 
\label{zetaeven}
\end{equation}
if $l+m$ is even, while
\begin{equation}
\zeta_{lm} = \frac {l+2}{l-1} \left [2 B_l\,  {_{-1}J}\, {_{-1}\, 
Y_{lm}} ({\textstyle \frac{\pi}{2}},0) \right]^2 
\label{zetaodd}
\end{equation}
if $l+m$ is odd. The symmetry $\eta_{l,-m} = \eta_{lm}$, used
in equation (\ref{power_1}), can easily be established from the
previous expressions.

Equations (\ref{power_1})--(\ref{zetaodd}) constitute the dressed
multipole truncation of the gravitational-wave luminosity. The
right-hand side of equation (\ref{power_1}) is an infinite series in 
powers of $v$, and our next task is to examine the convergence 
of this series.

\section{Convergence of dressed multipole truncation}
\label{3}

The velocity dependence of all the terms in equation (\ref{power_1}) is 
made clear by substituting equations (\ref{source}) and (\ref{AB}), 
which gives
\begin{equation}
\dot E = \dot E_Q \sum_{l=2}^{\infty} v^{2(l-2)} \sum_{m=1}^{l} \left \{ 
\begin{array}{ll} \left ( a_{lm} + b_{lm}v^2 +c_{lm} v^3 \right ) & l+m 
\mbox{~even} \\
v^2 \left ( d_{lm} + e_{lm} v^2 + f_{lm} v^3 \right ) & l+m \mbox{~odd}
\end{array} \right.  . \label{power_2}
\end{equation}
The various coefficients are
\begin{eqnarray}
a_{lm} &=&     \frac {5 \pi}{4} \frac {m^{2(l+1)}}{(2l+1)!!^2}
\frac {(l+1)(l+2)}{(l-1)l} \left [ Y_{lm} 
({\textstyle \frac{\pi}{2}},0) \right ]^2,  \label{alm} \\
b_{lm} &=&  -\left[ \frac {(l+9)m^2}{(l+1)(2l+3)} + 2l -1 \right]\, a_{lm},
\label{blm} \\
c_{lm} &=&     2 \pi m\, a_{lm}, \label{clm} \\
d_{lm} &=&     5 \pi\, \frac {m^{2(l+1)}}{(2l+1)!!^2} \frac
{l+2}{l-1} \left [ 
{_{-1}Y_{lm}} ({\textstyle \frac{\pi}{2}},0) \right ]^2,  \label{dlm} \\
e_{lm} &=&    -\left[ \frac {(l+4)m^2}{(l+2)(2l+3)} + \frac
{2(l-1)(l+2)}{l} - 3 \right]\, d_{lm}, \label{elm} \\
f_{lm} &=& 2 \pi m\, d_{lm}. \label{flm}  
\end{eqnarray}
We now express equation (\ref{power_2}) in the following form, which 
displays a more transparent grouping of terms:
\begin{equation}
\dot E = \dot E_Q \left \{ a_{22} + b_{22}v^2 + c_{22}v^3 +
\sum_{l=3}^{\infty} v^{2(l-2)} \Bigl[ S_1(v) + v^2 S_2(v) + v^3 S_3(v)
\Bigr] \right \},
\label{edot}
\end{equation}
where 
\begin{eqnarray}
S_1 (v) &=& \sum_{l=3}^{\infty} v^{2(l-2)} a_{ll}  
\left [ 1 + \frac{1}{a_{ll}} 
\sum_{m=1}^{l-1} \left ( a_{lm} + d_{l-1,m} \right ) \right ],
\label{S_1} \\
S_2(v) &=& \sum_{l=3}^{\infty} v^{2(l-2)} b_{ll}  
\left [ 1 + \frac{1}{b_{ll}}
\sum_{m=1}^{l-1} \left ( b_{lm} + e_{l-1,m} \right ) \right ],
\label{S_2}  \\
S_3(v) &=& \sum_{l=3}^{\infty} v^{2(l-2)} c_{ll}  
\left [ 1 + \frac{1}{c_{ll}} 
\sum_{m=1}^{l-1} \left ( c_{lm} + f_{l-1,m} \right ) \right ]. \label{S_3}  
\end{eqnarray}
The dressed multipole truncation therefore splits $\dot{E}$ into three 
separate series: $S_1(v)$, the bare multipole series, $v^2 S_2(v)$, the 
1PN correction, and $v^3 S_3(v)$, the 1.5PN correction. It is important 
to point out that in equations (\ref{S_1})--(\ref{S_3}), the sums over 
$m$ are restricted to values such that $l+m$ is {\it even}.

The convergence of the dressed multipole series may be established by
analyzing the convergence of each of the partial series. We begin with 
$S_1(v)$, the bare multipole series of reference \cite{paper_I}. Its 
convergence is established in the following manner. First, it is shown 
(numerically) that the terms in the square brackets of equation 
(\ref{S_1}) approach the approximate value 1.01 as 
$l \rightarrow \infty$. This 
implies that the behaviour of the series for large $l$ is dominated by 
the factor in front of the square brackets. The series may therefore be 
approximated by $S_1(v) \simeq \sum_{l=3}^{\infty} v^{2(l-2)} a_{ll}$, 
which is shown (using the Cauchy ratio test) to converge for
\begin{equation}
v < 2 / e \simeq 0.7358.  \label{v_range}
\end{equation}

The convergence of $S_2(v)$ is established in the same manner. 
First, one shows numerically that the terms in the square brackets of 
equation (\ref{S_2}) also approach 1.01 as $l \rightarrow \infty$. The 
series is then approximated by $S_2(v) \simeq \sum_{l=3}^{\infty} 
v^{2(l-2)} b_{ll}$, and the convergence of this series is again analyzed 
via the Cauchy ratio test. Using equation (\ref{blm}), we find that 
$b_{l+1,l+1}/b_{ll}$ is equal to $a_{l+1,l+1}/a_{ll}$ multiplied by a 
ratio of terms quadratic in $l$.  In the limit $l \rightarrow \infty$, 
$b_{l+1,l+1}/b_{ll} \sim a_{l+1,l+1}/a_{ll}$ and we immediately conclude
that the convergence of $S_2(v)$ is also determined by equation 
(\ref{v_range}).

The same is true also for $S_3(v)$.  Again, it can be shown that the 
terms in the square brackets of equation (\ref{S_3}) approach 1.01 as 
$l \rightarrow \infty$.  The series is then approximated by 
$S_3(v) \simeq \sum_{l=3}^{\infty} v^{2(l-2)} c_{ll}$.  
From equation (\ref{clm}) we see 
that in the limit $l \rightarrow \infty$, the ratio of successive 
coefficients is equal to $a_{l+1,l+1}/a_{ll}$. This allows us to 
conclude that the convergence of $S_3(v)$ is also determined by 
equation (\ref{v_range}).

In summary, when 1PN and 1.5PN corrections are added to the bare 
multipole series, $\dot{E}$ splits naturally into three separate 
series, which all converge for $v < 2/e$. Therefore, the dressed
multipole truncation of $\dot{E}$ must also converge in this interval.
We note that this convergence test ignores the fact that the combined
series $S(v) \equiv S_1(v)+v^2 S_2(v)+v^3 S_3(v)$ is alternating at 
large $l$. However, establishing the convergence of the alternating 
series using the Leibnitz criterion does not result in an improvement 
on equation (\ref{v_range}).  Furthermore, a numerical analysis of the 
convergence of $S(v)$ confirms equation (\ref{v_range}).

\section{Concluding remarks}
\label{4}

We have shown that for circular binaries with small mass
ratios, the dressed multipole truncation leads to a convergent series
for the gravitational-wave luminosity, for all physically relevant
values of the orbital velocity. It appears likely that this conclusion
will remain valid for binaries of comparable masses, but a separate
calculation is required to confirm this. Our results imply that the 
poor convergence of the true post-Newtonian series does not originate 
from summing over the Newtonian, 1PN, and 1.5PN part of all the 
multipole moments. 

We see in figure 1 that the dressed-multipole series leads to a 
significant improvement in accuracy, compared with the 
bare-multipole series.  By including the negative 1PN corrections 
and the (smaller) positive 1.5PN corrections, our approximation to 
the luminosity coincides (within 1 \%) with the numerically ``exact'' 
results \cite{Poisson} over the interval $v < 0.2$. One would 
therefore expect that the incorporation of further corrections 
would lead to a series that converges everywhere in the 
region of physical interest, and stays accurate over a larger interval. 

Meanwhile, the issue of the convergence of the true post-Newtonian
series remains an open one.

\section*{Acknowledgments}

This research was supported by the Natural Sciences and Engineering
Research Council of Canada. Conversations with Clifford Will were
greatly appreciated.

\newpage

\begin{figure}

\special{hscale=50 vscale=50 hoffset=20.0 voffset=50.0
         angle=-90 psfile=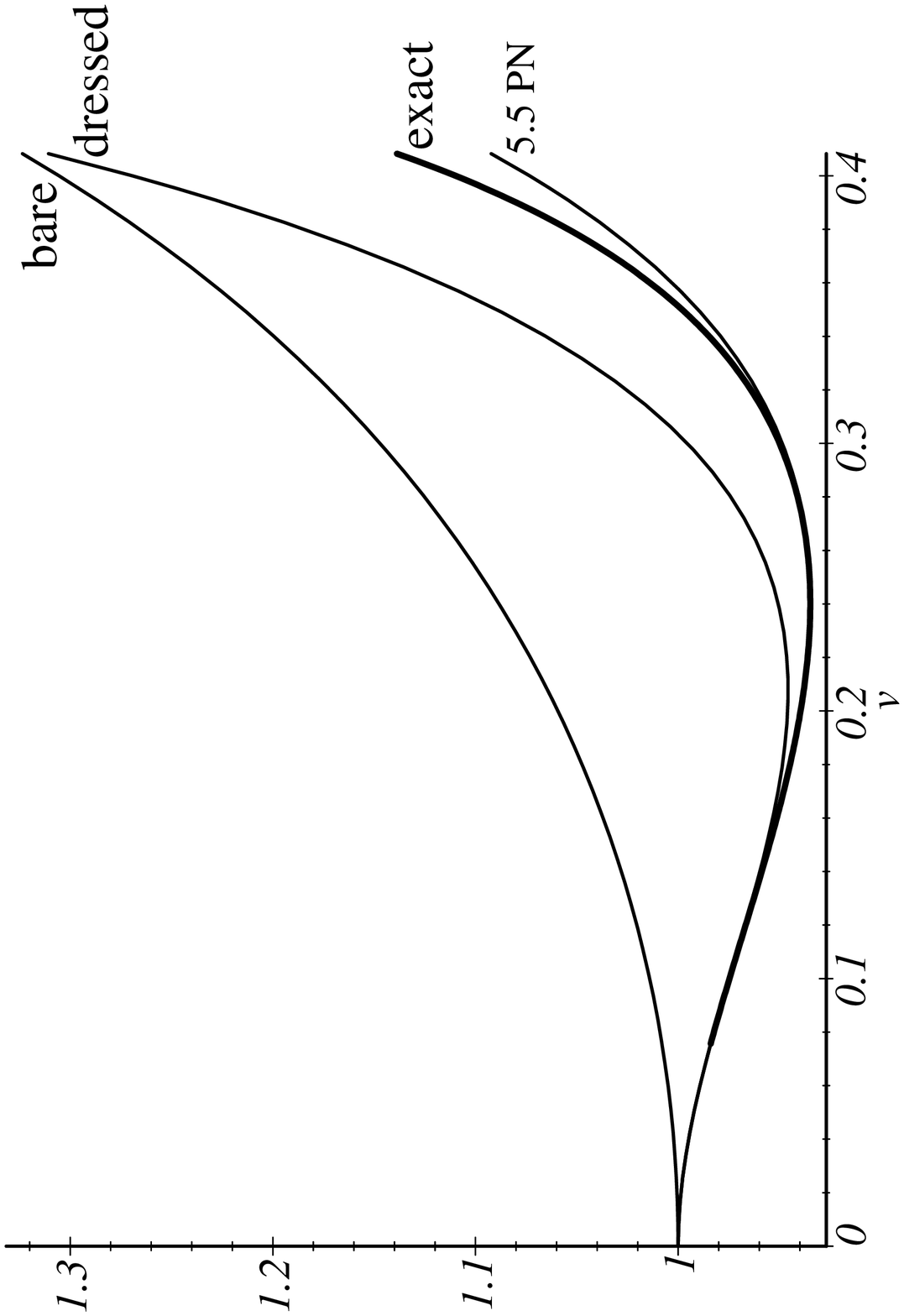}

\vspace*{4in}
\caption[plot]{Plots of $\dot{E}/\dot{E}_Q$ as a function of orbital 
velocity $v$. The thick curve is the ``exact'' curve, obtained by 
numerically integrating the equations of black-hole perturbation 
theory (see reference \cite{Poisson}). The curve immediately below the 
exact curve is the 5.5PN curve, obtained from the post-Newtonian 
expansion (carried up to order $v^{11}$ beyond the quadrupole formula) 
of Tanaka {\it et al.}~\cite{Tagoshietal}. The first curve above the 
exact curve is the dressed multipole truncation, obtained from 
equation (\ref{edot}). Finally, the highest curve is the bare multipole 
truncation, obtained also from equation (\ref{edot}), but with $S_2(v)$ 
and $S_3(v)$ set to zero. We see that the accuracy of the dressed
multipole truncation is far superior to that of the bare multipole
truncation, but that it does not match the accuracy of the 5.5PN
approximation.}

\end{figure}
\end{document}